\documentclass{article}
\usepackage{amsmath,graphicx,mlspconf}
\usepackage{array}
\usepackage{multirow}
\usepackage{booktabs}
\usepackage{caption}
\usepackage{makecell}
\usepackage{cite}
\usepackage{amsmath,amssymb,amsfonts}
\usepackage{algorithmicx}
\usepackage{algorithm}
\usepackage{algpseudocode}
\usepackage{graphicx}
\usepackage{textcomp}
\usepackage{color}
\usepackage{xcolor}
\usepackage{url}
 
\usepackage{etoolbox}

\patchcmd{\thebibliography}
  {\settowidth}
  {\setlength{\itemsep}{-2pt plus 0.01ex}%
   \setlength{\parskip}{-1pt}%
   \settowidth}
  {}{}

\toappear{Preprint submitted to IEEE MLSP 2025}

\title{Ptychographic Image Reconstruction from Limited Data via Score-Based Diffusion Models with Physics-Guidance}
\name{Refik M. Cam$^1$\sthanks{The author conducted this work while at Argonne National Laboratory.}, Junjing Deng$^2$,  Rajkumar Kettimuthu$^2$, Mathew J. Cherukara$^2$, Tekin Bicer$^2$}
\address{$^1$University of Illinois Urbana-Champaign, IL 61801, USA
\\$^2$Argonne National Laboratory, IL 60439, USA}

\begin{document}
%\ninept

\maketitle

\begin{abstract}
Ptychography is a data-intensive computational imaging technique that achieves high spatial resolution over large fields of view. The technique involves scanning a coherent beam across overlapping regions and recording diffraction patterns. Conventional reconstruction algorithms require substantial overlap, increasing data volume and experimental time, reaching PiB-scale experimental data and weeks to month-long data acquisition times. To address this, we propose a reconstruction method employing a \textit{physics-guided score-based diffusion model}. Our approach trains a diffusion model on representative object images to learn an object distribution prior. During reconstruction, we modify the reverse diffusion process to enforce data consistency, guiding reverse diffusion toward a physically plausible solution. This method requires a single pretraining phase, allowing it to generalize across varying scan overlap ratios and positions. Our results demonstrate that the proposed method achieves high-fidelity reconstructions with only a 20\% overlap, while the widely employed rPIE method requires a 62\% overlap to achieve similar accuracy. This represents a significant reduction in data requirements, offering an alternative to conventional techniques.
\end{abstract}

\section{Introduction}
\label{sec:intro}
Ptychography has emerged as a versatile imaging tool across various scientific fields, including materials science, biology, and nanotechnology \cite{pfeiffer2018x}. X-ray ptychography, widely adopted at synchrotron radiation facilities \cite{APS_comp_strategy}, enables detailed examination of large volumes in thick samples with minimal preparation, benefiting applications such as imaging integrated circuits (ICs), biological specimens, and strain analysis in nanostructures \cite{pfeiffer2018x,benmore2022advancing, zhao2024suppressing}.

\begin{figure}[h]
    \centering
    \vspace{-15pt}
    \includegraphics[width=0.7\linewidth]{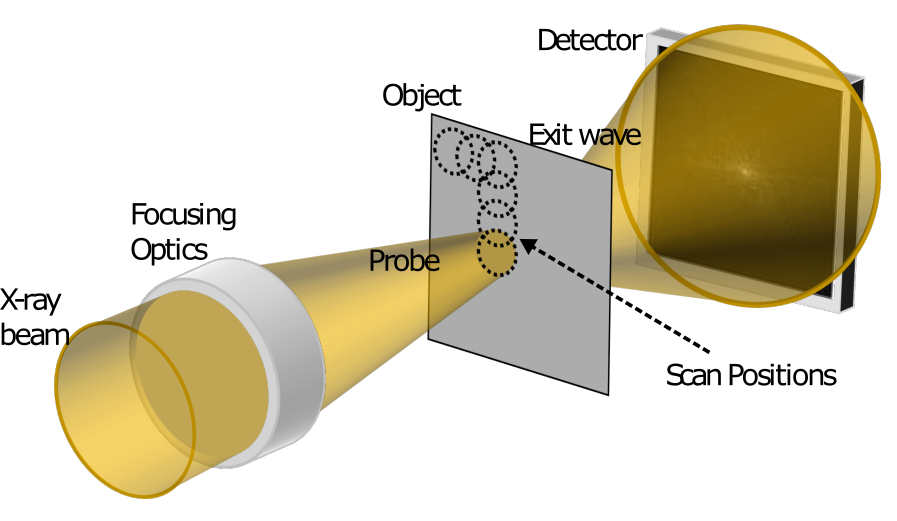}
    \vspace{-8pt}
    \caption{Schematic of the ptychographic imaging process.}
    \label{ptychographic_imaging}
\vspace{-15pt}
\end{figure}

The ptychographic imaging process (Fig. \ref{ptychographic_imaging}) involves scanning a coherent beam across overlapping regions of a sample and recording far-field diffraction patterns. The sample's complex-valued transmission function is then computationally reconstructed from these intensity-only measurements. This reconstruction process is inherently ill-posed, as phase information is lost in the recorded diffraction patterns. Traditional reconstruction methods require substantial overlap between probe positions to mitigate ill-posedness, which increases data volume, acquisition time, and computational requirements \cite{babu2023deep, yu2022scalable, yu2021topology}. For instance, imaging a 1~cm$^2$ IC at sub-10~nm resolution using the Microscopy beamlines at Advanced Photon Source (APS) results in PiB-scale dataset, requires close to a month-long data acqusition and supercomputer-scale resources for timely reconstruction \cite{aps-raven}. Such experimental dataset and compute requirements do not only limit the advanced experimentation at synchrotron radiation facilities, but also hinder the science relying on them. %These constraints limit throughput and expose dose-sensitive samples to prolonged radiation, risking damage. %\cite{guan2019ptychonet, yu2022scalable}.

\vspace{-5pt}

To address these challenges, we develop a ptychographic image reconstruction framework using \textit{score-based diffusion models}. Our method trains a diffusion model on representative object images, learning an object distribution as a prior. During reconstruction, we adapt the reverse diffusion process to enforce consistency with measured diffraction patterns, guiding the reverse diffusion toward a physically plausible reconstruction. By significantly reducing the required overlap ratio between measurements, our approach enables faster acquisitions, reduced data storage and computation, and better preservation of dose-sensitive samples while providing accurate and high-quality reconstruction. To the best of our knowledge, this work is the first to develop, apply, and evaluate diffusion models for real-space ptychographic image reconstruction.\footnote{The code, trained model, and the sample dataset can be accessed at [https://github.com/hidden-due-to-double-blind-review]}%https://github.com/refik-mert-cam/ptychography_diffusion_model_based_reconstruction.git }. %making it suitable for high-throughput and dose-sensitive imaging applications.

\section{Background}
\vspace{-5pt}
\subsection{Ptychographic Imaging Inverse Problem}
\vspace{-8pt}
Ptychographic image reconstruction aims to estimate an object \( \mathbf{f} \in \mathbb{C}^n \) from measurements \( \{ \mathbf{y}_i \in \mathbb{R}^m \}_{i=1}^K \) expressed as \cite{gan2024ptychodv, zhai2023projected, xu2018accelerated}:
\begin{equation}
\mathbf{y}_i^2 \sim \text{Pois}(| \mathbf{F} \mathbf{P} \mathbf{D}_i \mathbf{f} |^2),
\label{eq:ptycho}
\end{equation}
where \( \mathbf{P} \in \mathbb{C}^{m \times m} \) is the complex-valued probe illumination, \( \mathbf{F} \in \mathbb{C}^{m \times m} \) is the Fourier transform, and \( \text{Pois}(\cdot) \) denotes the Poisson distribution, reflecting the quantum nature of photon detection. The matrix \( \mathbf{D}_i \in \{0, 1\}^{m \times n} \) extracts a patch from the image \( \mathbf{f} \), corresponding to the \( i \)-th probe position. \( K \) is the total number of probe positions. 
Synchrotron radiation facilities, such as APS, require large overlap between consecutive measurements \( \mathbf{y}_i \)---typically above 65\%---to ensure accurate reconstructions, which, depending on the size of the object (view) and target resolution, can result in PiB-scale measurement dataset per imaged sample \cite{aps-raven}. 
%While some approaches jointly estimate the object \( \mathbf{f} \), the probe function \( \mathbf{P} \), and the scan locations \( \{ \mathbf{D}_i \} \) \cite{maiden2017further}; in this study, the probe function \( \mathbf{P} \) and scan locations \( \{ \mathbf{D}_i \} \) are assumed to be known, and only the object \( \mathbf{f} \) is estimated.

\vspace{-10pt}
\subsection{Ptychographic Image Reconstruction}
\vspace{-8pt}
Iterative algorithms have long been used for ptychographic image reconstruction, but they often fail under low probe overlap conditions \cite{maiden2017further, candes2015phase, xu2018accelerated}. To overcome this, deep learning (DL) methods such as PtychoNet \cite{guan2019ptychonet} and PtychoNN \cite{cherukara2020ai} have been proposed to directly map individual diffraction patterns to object patches. However, these models neglect the inherent redundancy from overlapping scans, leading to hallucinations and limited accuracy \cite{gan2024ptychodv}. To address this, PtychoDV \cite{gan2024ptychodv} was introduced, combining a vision transformer for initial estimation and an unrolled optimization network with learnable convolutional priors for refinement. While effective, it requires retraining for each probe overlap ratio. Generative models, particularly generative adversarial networks (GANs), have also been applied to improve reconstructions and uncertainty quantification under low-overlap conditions \cite{barutcu2022compressive, ekmekci2024integrating}, though they are hindered by mode collapse and training instability. Recently, diffusion models have surpassed GANs in fidelity and robustness across multiple domains \cite{dhariwal2021diffusion}, yet remain unexplored in the context of real-space ptychography.

\vspace{-10pt}
\subsection{Score-Based Diffusion Models}
\vspace{-8pt}
Diffusion models provide a powerful generative framework that samples from complex data distributions by reversing a process of incremental noise addition. In the forward process, data are gradually transformed into a simpler---typically Gaussian---distribution. This transformation is characterized by the stochastic differential equation (SDE) \cite{chung2022diffusion}:
\begin{equation}
d\mathbf{x_t} = -\frac{1}{2} \beta(t) \mathbf{x_t} \, dt + \sqrt{\beta(t)} \, d\mathbf{w}, \quad t \in [0, T],
\end{equation}
where \( \mathbf{x_t} \) represents the state at time \( t \), \( \beta(t) \) is a time-dependent noise schedule, and \( d\mathbf{w} \) denotes the standard Wiener process.

To retrieve samples from the target data distribution, the forward noising process is inverted by solving the reverse-time SDE:
\begin{equation}
d\mathbf{x_t} = \left[ -\frac{\beta(t)}{2}  \mathbf{x_t} - \beta(t) \nabla_{\mathbf{x_t}} \log p_t(\mathbf{x_t}) \right] \, dt + \sqrt{\beta(t)} \, d\overline{\mathbf{w}},
\label{eq:reverse-sde-posterior}
\end{equation}
where \( \nabla_{\mathbf{x_t}} \log p_t(\mathbf{x_t}) \) is the score function, representing the gradient of the log-probability with respect to the noisy data. Since the true score is intractable, a neural network is trained to approximate it at each time step. Once the diffusion model is trained, it can serve as data-driven prior to regularizing ill-posed inverse problems, as it implicitly captures the underlying object distribution.

% The exact score function is typically unknown and is approximated using a neural network. 

\vspace{-10pt}
\subsection{Diffusion Posterior Sampling}
\vspace{-8pt}
To solve inverse problems, diffusion models can be adapted to sample from the posterior distribution \( p(\mathbf{x}|\mathbf{y}) \) by introducing a likelihood-based guidance step into the reverse-time SDE. The modified reverse-time SDE is expressed as \cite{chung2022diffusion}:
\begin{align}
d\mathbf{x}_t = &\left[ -\frac{\beta(t)}{2} \mathbf{x}_t - \beta(t) (\nabla_{\mathbf{x}_t} \log p_t(\mathbf{x}_t) \right. \nonumber \\
&\left. + \nabla_{\mathbf{x}_t} \log p_t(\mathbf{y}|\mathbf{x}_t)) \right] \, dt + \sqrt{\beta(t)} \, d\overline{\mathbf{w}}.
\label{eq:reverse-sde-posterior}
\end{align}
where the likelihood gradient 
$\nabla_{\mathbf{x}_t} \log p_t(\mathbf{y}|\mathbf{x}_t)$ 
ensures data consistency. This term can be approximated by evaluating \( \nabla_{\mathbf{x}_t}\log p_t(\mathbf{y}|\mathbf{\hat{x}_0}) \), where \( \mathbf{\hat{x}_0} \) is the expected initial state given the current state \( \mathbf{x_t} \), i.e., \( \mathbf{\hat{x}_0} := \mathbb{E}[\mathbf{x_0} | \mathbf{x_t}] \) \cite{chung2022diffusion}. For variance-preserving SDEs or Denoising Diffusion Probabilistic Model (DDPM) sampling, a closed-form approximation simplifies gradient computation \cite{chung2022diffusion}. Diffusion posterior sampling is well-suited for non-linear inverse problems as it integrates learned object priors with gradient-based data consistency.

\begin{figure*}[h!]
    \centering
    \vspace{-15pt}
    \includegraphics[width=\textwidth]{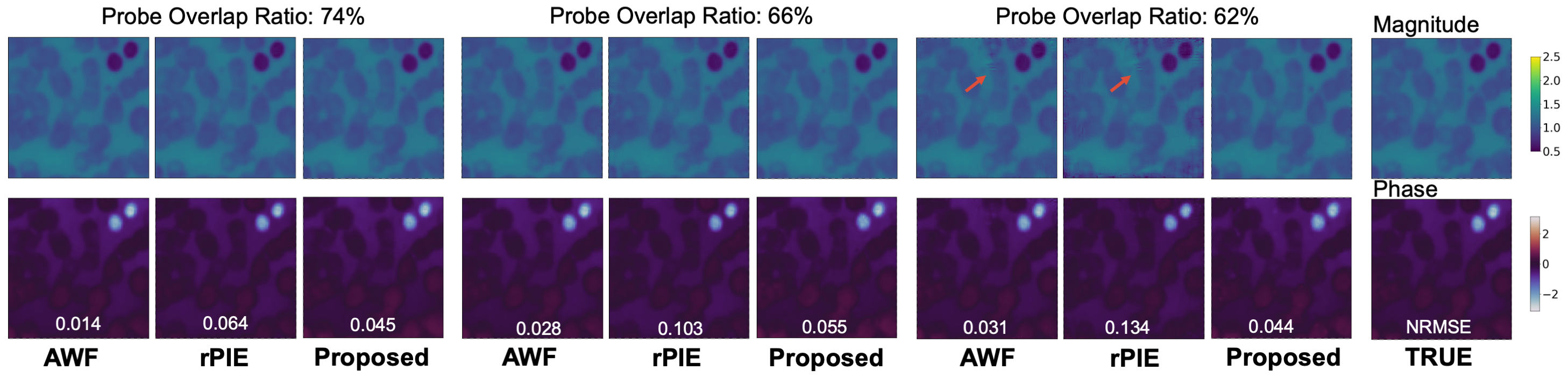}
    \caption{Sample reconstructions for high probe overlap ratios (74\%, 66\%, 62\%). The proposed method consistently produces accurate reconstructions, even as conventional iterative methods begin to introduce artifacts at the overlap ratio of 62\%.}
    \label{sample_reconstructions_high_overlap}
\vspace{-10pt}
\end{figure*}

\section{Method}
\vspace{-10pt}
\begin{algorithm}
\caption{Diffusion Model-Based Ptychographic Image Reconstruction with Time-Travel Strategy}
{\fontsize{9pt}{11pt}\selectfont % Set font size to 9pt with a 11pt line spacing
\begin{algorithmic}[1]
\Require $N$, $\mathbf{y}$, $j$, $\zeta$, $\{\bar{\alpha}_t\}_{t=1}^N$, $\{\beta_t\}_{t=1}^N$, $\{\tilde{\sigma}_t\}_{t=1}^N$
%\State $\mathbf{x}_N \sim \mathcal{N}(\mathbf{0}, \mathbf{I})$ \Comment{Initialization with Gaussian noise}
\For{$t = N-1$ \textbf{down to} $0$} 
    \State $\mathbf{x}_{t-1} \gets \text{DenoisingStep}(\mathbf{x}_{t}, \bar{\alpha}_t, \beta_t, \tilde{\sigma}_t)$
    \State $\mathbf{x}_{t-1} \gets \mathbf{x}_{t-1} - \zeta \nabla_{\mathbf{x}_t} \| \mathbf{y} - |\mathbf{A}\hat{\mathbf{x}}_0| \|_1$ \Comment{Update using gradient of the energy function (likelihood substitute) with $\ell 1$-norm}
    \If{$t + j \leq N-1$} \Comment{Time-travel strategy}
        \State $\mathbf{x}_{t+j} \gets \text{NoisingStep}(\mathbf{x}_{t}, j)$ \Comment{Revert state by $j$ steps}
        \For{$i = 0$ \textbf{to} $j-1$}
            \State $\mathbf{x}_{t+j-i-1} \gets \text{DenoisingStep}(\mathbf{x}_{t+j-i}, \bar{\alpha}_{t+j-i}, ... \newline \beta_{t+j-i}, \tilde{\sigma}_{t+j-i})$
            \State $\mathbf{x}_{t+j-i-1} \gets \mathbf{x}_{t+j-i-1} - \zeta \nabla_{\mathbf{x}_{t+j-i}} \| \mathbf{y} - |\mathbf{A}\hat{\mathbf{x}}_0| \|_1$
        \EndFor
    \EndIf
\EndFor
\State \Return $\hat{\mathbf{x}}_0$
\end{algorithmic}
\par} % End the scope of the font size setting
\label{ptycho_algorithm}
\end{algorithm}
\vspace{-10pt}

We implemented a denoising diffusion probabilistic model with a U-Net backbone conditioned on diffusion process time steps. The network is configured to process two-channel inputs and outputs, where the channels explicitly represent the amplitude and phase of the complex-valued object. At each diffusion step, the model receives a noisy two-channel input and produces a denoised two-channel output, preserving the distinction between amplitude and phase throughout the reverse process. This design is well aligned with the complex-valued nature of ptychographic imaging and is particularly effective due to the inherent correlation between the amplitude and phase components. 

% The U-Net backbone of the employed diffusion model follows a five-level encoder-decoder architecture, with attention mechanisms applied at spatial resolutions of 32, 16, and 8 to capture the long-range dependencies in the object. 
The U-Net backbone of the employed diffusion model adopts a five-level encoder–decoder architecture, with attention mechanisms integrated at the third, fourth, and fifth levels to capture the long-range dependencies in the object representation. Each level contains two residual blocks, each consisting of convolutional layers with a $3 \times 3$ kernel, starting with a base channel width of 128 and progressively expanding based on spatial scale. To enhance non-local feature learning, we apply multi-head self-attention with 4 heads and 64 channels per head in the designated layers. The model is also conditioned on the current diffusion time step via learnable embeddings that modulate intermediate feature activations, allowing the network to adapt its denoising behavior as the reverse process progresses. These design choices, including the placement of attention layers and number of layers, were heuristically determined based on empirical performance during preliminary experimentation.

Although the ptychography experiments result in a massive number of measurements (\( \mathbf{y}_i \) in Eq. \ref{eq:ptycho}), the number of real-space object views is typically very limited (\( \mathbf{f} \) in Eq. \ref{eq:ptycho}) %Ptychographic imaging experiments typically produce only a limited number of high-quality samples
---often on the order of tens of images from sub-sections of sample view or different rotations---due to constraints in acquisition time, sample availability, or experimental complexity. This data scarcity presents a significant challenge for training data-hungry generative models such as diffusion models. To address this, we implement extensive data augmentation strategies that are physically consistent with the ptychographic imaging process. In particular, random rotations, scaling, and cropping are natural augmentations, as they reflect the inherent variability in sample orientation, magnification, and field-of-view in experimental setups and data acquisition. These transformations preserve the underlying physical structures while providing the necessary diversity to train a robust model.

The proposed ptychographic image reconstruction method \textit{leverages the trained diffusion model as an object prior} and builds upon the diffusion posterior sampling framework, where data consistency is enforced by incorporating an approximate log-likelihood term into the reverse-time SDE. Specifically, we employ the following:
\[
\log p_t(\mathbf{y}|\hat{\mathbf{x}}_0) \propto \left\| \mathbf{y} - \left| \mathbf{A} \hat{\mathbf{x}}_0 \right| \right\|_1.
\]
While both $\ell_1$ and $\ell_2$ norms were evaluated in this context, we found that the $\ell_1$-norm led to more robust and accurate reconstructions. This improvement is mainly due to the inherent sparsity of the diffraction measurements, for which $\ell_1$-based losses are better suited. The forward model $\mathbf{A}$ is composed of a sequence of operators---object patch extraction, probe modulation, and Fourier transform---represented as $\{\mathbf{F} \mathbf{P} \mathbf{D}_i\}_{i=1}^K$, applied across multiple scanning positions. Gradients of the log-likelihood term with respect to the image estimate $\hat{\mathbf{x}}_0$ are computed using automatic differentiation provided by PyTorch, enabling seamless integration into the sampling dynamics.

A limitation of the diffusion posterior sampling is the potential for insufficient likelihood guidance, which can lead to suboptimal reconstructions. To mitigate this, we introduce an enhancement where additional sampling steps are employed to boost sample diversity and prevent premature convergence, which has proven to be effective in other applications \cite{lugmayr2022repaint}. Specifically, this strategy involves perturbing the state backward by \( j \) steps and then re-denoising, refining intermediate reconstructions. Through this controlled resampling, the reconstruction is steered toward a solution that better aligns with the measurements. The corresponding algorithm is detailed in Algorithm \ref{ptycho_algorithm}, where $\{\bar{\alpha}_t\}_{t=1}^N$, $\{\beta_t\}_{t=1}^N$, and $\{\tilde{\sigma}_t\}_{t=1}^N$ are predefined parameters for forward and reverse SDEs, and \( \zeta \) represents the step-size. We found that \( N = 1000\) total steps and \( j = 10 \) backward steps yield accurate ptychographic reconstructions in our experiments.

% \textbf{Dataset and training:} 

% To simulate this limited-data setting and demonstrate the feasibility of applying diffusion models to ptychographic imaging, we constructed a dataset using four high-resolution ($512 \times 512$) Fourier ptychographic images of blood smears \cite{zheng2021concept}. From these, we generated 100,000 training samples by randomly rotating, scaling, and cropping the three full images and the left half ($512 \times 256$) of the fourth image into $128 \times 128$ patches. The test set, consisting of 50 images, was created from the right half of the fourth image to ensure no spatial overlap with the training data. During training, both the amplitude and phase components of the complex-valued images are linearly scaled to the range $[-1, 1]$ to improve numerical stability and facilitate gradient-based learning. This normalization ensures that the input and output distributions are centered and bounded, which is particularly beneficial for stable diffusion-based denoising. After the reverse diffusion process, the model's output is rescaled back to its original dynamic range, as determined from the empirical minimum and maximum values observed in the training set. The model was trained using the mean squared error (MSE) loss between the predicted and true noise at each diffusion step. Training was performed over 130,000 steps with a batch size of 8 and a constant learning rate of $10^{-4}$. 

\begin{figure*}[h!]
    \vspace{-10pt}
    \centering
    \includegraphics[width=0.8\textwidth]{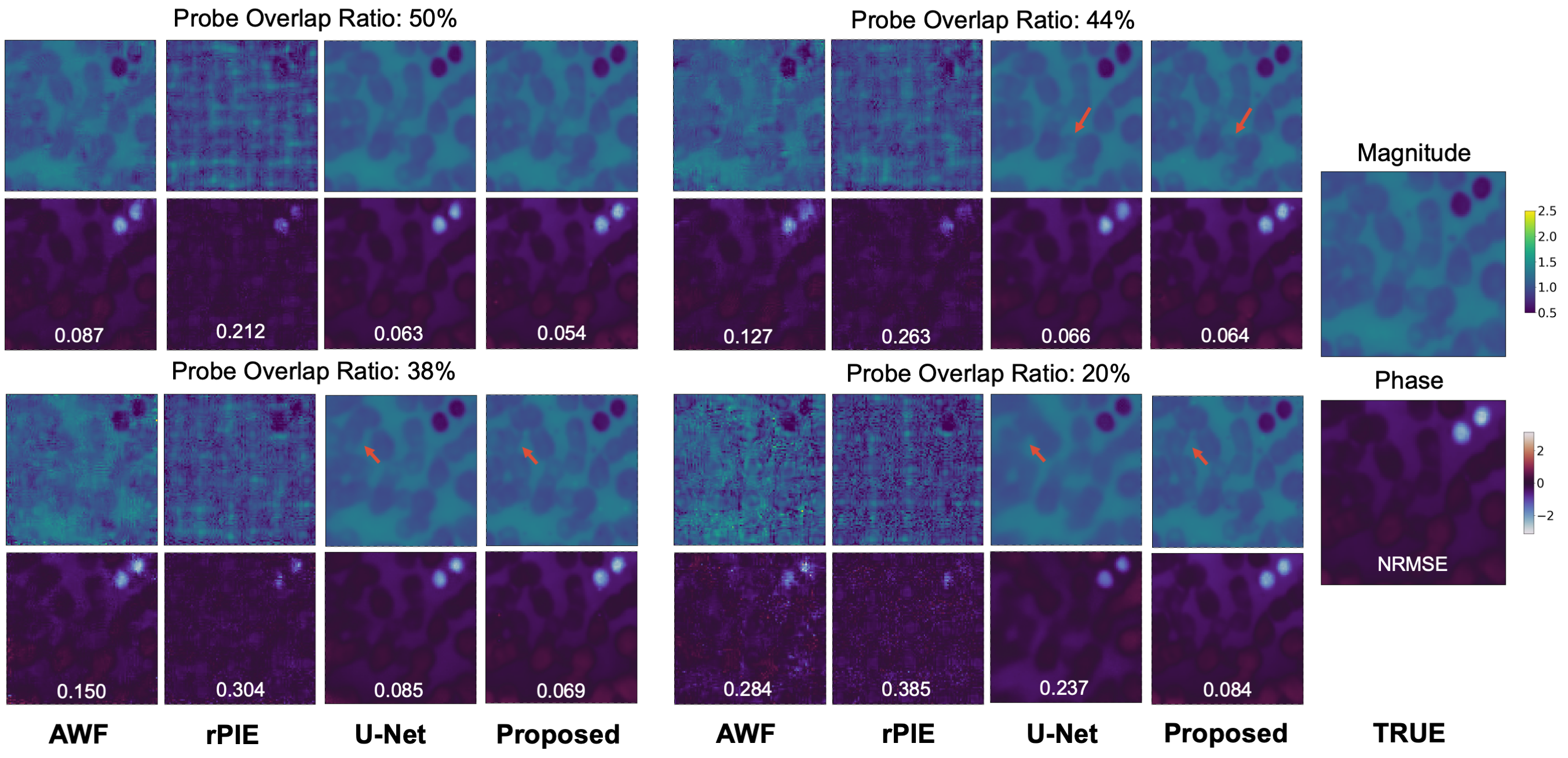}
    \vspace{-5pt}
    \caption{Sample reconstructions for low probe overlap ratios. The comparison between different methods is shown for overlap ratios of 50\%, 44\%, 38\%, and 20\%. The proposed method consistently produces accurate reconstructions, outperforming the baseline methods as the overlap ratio decreases.}
    \label{sample_reconstructions_low_overlap}
\vspace{-15pt}
\end{figure*}

\section{Experiments}
\vspace{-10pt}
We evaluated the proposed method on a dataset derived from four high-resolution ($512 \times 512$) Fourier ptychographic images of blood smears \cite{zheng2021concept}. As discussed in the Methods section, this setup simulates the data-scarce conditions typical of experimental ptychography. To address this, we applied extensive data augmentations---random rotations, scaling, and cropping---to generate 100,000 training patches of size $128 \times 128$ from three full images and the left half ($512 \times 256$) of the fourth image. The test set comprised 50 patches sampled from the right half of the fourth image, ensuring that the proposed reconstruction method was evaluated on a spatial region entirely unseen during training. Complex-valued images were normalized to the range $[-1, 1]$ during training to stabilize the diffusion process and rescaled to their original dynamic range after sampling. The diffusion model was trained for 130,000 steps using mean squared error (MSE) loss between predicted and true noise, with a batch size of 8 and a constant learning rate of $10^{-4}$. The resulting model was then used as the object prior in the proposed sampling-based reconstruction pipeline.

% Numerical experiments were conducted to evaluate the performance of the proposed method. To simulate this limited-data setting and demonstrate the feasibility of applying diffusion models to ptychographic imaging, we constructed a dataset using four high-resolution ($512 \times 512$) Fourier ptychographic images of blood smears \cite{zheng2021concept}. From these, we generated 100,000 training samples by randomly rotating, scaling, and cropping the three full images and the left half ($512 \times 256$) of the fourth image into $128 \times 128$ patches. The test set, consisting of 50 images, was created from the right half of the fourth image to ensure no spatial overlap with the training data. During training, both the amplitude and phase components of the complex-valued images are linearly scaled to the range $[-1, 1]$ to improve numerical stability and facilitate gradient-based learning. This normalization ensures that the input and output distributions are centered and bounded, which is particularly beneficial for stable diffusion-based denoising. After the reverse diffusion process, the model's output is rescaled back to its original dynamic range, as determined from the empirical minimum and maximum values observed in the training set. The model was trained using the mean squared error (MSE) loss between the predicted and true noise at each diffusion step. Training was performed over 130,000 steps with a batch size of 8 and a constant learning rate of $10^{-4}$. This trained diffusion model was employed as the object prior. 

A ptychographic imaging system was simulated using a 32$\times$32 pixel probe, downsampled from the probe used in \cite{zhai2023projected}. To assess the method under varying imaging conditions, different probe overlap ratios---20\%, 38\%, 44\%, 50\%, 62\%, 66\%, and 74\%---were simulated on 128$\times$128 pixel objects. Pseudo-Poisson noise was introduced into the diffraction patterns, assuming a maximum photon count of 10$^5$.

To establish a baseline for performance evaluation, three reconstruction methods were implemented: the rPIE \cite{maiden2017further}, Accelerated Wirtinger Flow (AWF) \cite{xu2018accelerated}, and a U-Net-based post-processing approach \cite{ronneberger2015u}. The rPIE and AWF methods were executed for 10,000 iterations to ensure convergence. For cases where the probe overlap ratio was at or below 50\%, the U-Net-based post-processing technique was employed in addition to rPIE and AWF. This technique was included as it represents a commonly used post-processing strategy in image reconstruction pipelines \cite{antholzer2019deep}. However, it should be noted that the U-Net must be retrained for each specific imaging configuration, as its performance depends heavily on the statistical characteristics of the data.

The U-Net-based post-processing aimed to enhance AWF reconstructions by mitigating artifacts and improving overall image quality. It processes two-channel input images (amplitude and phase) and outputs two-channel reconstructions. It employs a five-level encoder-decoder architecture with residual blocks and skip connections, starting with 32 channels and progressively increasing to 512. The design parameters were heuristically determined based on empirical performance. The U-Net was trained using MSE loss utilizing 20,000 AWF reconstructions, with 19,000 used for training and 1,000 for validation. Training was conducted over 800 epochs using the ADAM optimizer with an initial learning rate of $2\times10^{-5}$, and the model with the lowest validation loss was selected. Data augmentation included random flipping and Gaussian noise injection with a standard deviation of 0.01 in half of the training batches.

Normalized root-mean-squared error (NRMSE) and structural similarity index (SSIM) were used to evaluate the quality of the reconstructed images. To address phase ambiguity, a phase shift was incorporated into the NRMSE calculation \cite{zhai2023projected}. SSIM was computed over the magnitude images.

\vspace{-5pt}
\begin{table}[h]
\centering
\caption{Reconstruction accuracy of different methods. The first row in each cell represents NRMSE, and the second row represents SSIM.}
\label{reconstruction_accuracy_table}
\vspace{-5pt}
\resizebox{\linewidth}{!}{
\begin{tabular}{c|cccc}
\toprule
 \makecell{Overlap\\ratio} & rPIE & AWF & U-Net & Proposed \\ \midrule
 20\% & \makecell{0.388$\pm$0.019 \\ 0.146$\pm$0.015} & \makecell{0.283$\pm$0.022 \\ 0.228$\pm$0.025} & \makecell{0.135$\pm$0.026 \\ 0.854$\pm$0.027} & \makecell{\textbf{0.116$\pm$0.048} \\ \textbf{0.899$\pm$0.070}} \\ \midrule
 38\% & \makecell{0.309$\pm$0.023 \\ 0.284$\pm$0.020} & \makecell{0.153$\pm$0.020 \\ 0.572$\pm$0.051} & \makecell{0.085$\pm$0.017 \\ 0.927$\pm$0.023} & \makecell{\textbf{0.079$\pm$0.022} \\ \textbf{0.946$\pm$0.025}} \\ \midrule
44\% & \makecell{0.260$\pm$0.028 \\ 0.390$\pm$0.027} & \makecell{0.111$\pm$0.019 \\ 0.714$\pm$0.056} & \makecell{0.078$\pm$0.020 \\ 0.947$\pm$0.022} & \makecell{\textbf{0.071$\pm$0.021} \\ \textbf{0.957$\pm$0.020}} \\ \midrule
50\% & \makecell{0.211$\pm$0.035 \\ 0.562$\pm$0.034} & \makecell{0.082$\pm$0.015 \\ 0.832$\pm$0.046} & \makecell{0.067$\pm$0.020 \\ 0.947$\pm$0.022} & \makecell{\textbf{0.062$\pm$0.013} \\ \textbf{0.967$\pm$0.015}} \\ \midrule 
62\% & \makecell{0.140$\pm$0.038 \\ 0.897$\pm$0.022} & \makecell{\textbf{0.040$\pm$0.012} \\ 0.972$\pm$0.015} & \makecell{N/A \\ N/A} & \makecell{0.054$\pm$0.013 \\ \textbf{0.977$\pm$0.009}} \\ \midrule
66\% & \makecell{0.104$\pm$0.043 \\ 0.976$\pm$0.012} & \makecell{\textbf{0.026$\pm$0.011} \\ \textbf{0.991$\pm$0.009}} & \makecell{N/A \\ N/A} & \makecell{0.049$\pm$0.011 \\ 0.985$\pm$0.004} \\ \midrule
 74\% & \makecell{0.058$\pm$0.026 \\ 0.997$\pm$0.002} & \makecell{\textbf{0.019$\pm$0.009} \\ \textbf{0.999$\pm$0.001}} & \makecell{N/A \\ N/A} & \makecell{0.046$\pm$0.014 \\ 0.990$\pm$0.003} \\ \bottomrule
\end{tabular}
}
\vspace{-15pt}
\end{table}

\vspace{-5pt}
\subsection{Image reconstruction quality}
\vspace{-5pt}
In Fig. \ref{sample_reconstructions_high_overlap}, sample reconstructions are shown for probe overlap ratios of 74\%, 66\%, and 62\%. At these overlap ratios, all methods produce visually similar, accurate reconstructions. However, as the overlap decreases to 62\%, AWF and rPIE begin to introduce artifacts (marked with red arrows). For lower overlap ratios (50\%, 44\%, 38\%, and 20\%), the sample reconstructions are shown in Fig. \ref{sample_reconstructions_low_overlap}. The U-Net post-processing method effectively mitigates artifacts in AWF reconstructions, achieving results visually comparable to those of the proposed method. However, closer inspection reveals that U-Net struggles to recover certain structures (marked with red arrows).

Table \ref{reconstruction_accuracy_table} presents the reconstruction accuracy of different methods across various probe overlap ratios, measured by NRMSE and SSIM computed over the test set of 50 images. As the probe overlap ratio increases, the reconstruction quality improves significantly for all methods. At lower probe overlap ratios (20\% to 50\%), the proposed method consistently outperforms others, achieving the lowest NRMSE and highest SSIM values, indicating superior accuracy and structural preservation. While the U-Net-based post-processing approach also yields substantial improvements over traditional iterative methods, it is important to note that this technique requires retraining for each distinct imaging setting to maintain effectiveness.

% Notably, at 20\% overlap, the proposed method achieves an SSIM of 0.899, significantly higher than the second-best method at 0.854.

\vspace{-10pt}
\subsection{Generalization with Varied Probe Positions}
\vspace{-5pt}
The proposed method requires only a single pretraining phase on object distributions and exhibits robust generalization across varying scan overlap ratios and probe positions. This adaptability is important in practical applications, where scan parameters may fluctuate due to experimental constraints. To assess this generalization capability, we simulated a scenario in which the test set employed 20\% probe overlap ratio, like previously considered in this study, but with different probe positions. In this new configuration, the proposed method demonstrated minimal performance variation (NRMSE: 0.116 $\pm$ 0.048 $\rightarrow$ 0.117 $\pm$ 0.049, SSIM: 0.899 $\pm$ 0.070 $\rightarrow$ 0.902 $\pm$ 0.065), whereas the U-Net post-processing approach exhibited significant degradation (NRMSE: 0.135 $\pm$ 0.026 $\rightarrow$ 0.157 $\pm$ 0.036, SSIM: 0.854 $\pm$ 0.027 $\rightarrow$ 0.838 $\pm$ 0.030). These values represent the mean and standard deviation evaluated on a test set of 50 images. Furthermore, as illustrated in Table~\ref{tab:generalization}, the performance of the U-Net post-processing method deteriorates markedly when tested on probe overlap ratio settings different from those it was trained on, underscoring its limited generalization. This limitation requires retraining for different measurement settings, whereas the proposed method eliminates the need for retraining.

\vspace{-3pt}
\begin{table}[h]
\centering
\vspace{-5pt}
\caption{Generalization performance of the U-Net post-processing image reconstruction approach}
\label{tab:generalization}
\vspace{-5pt}
\resizebox{\linewidth}{!}{
\begin{tabular}{cc|cccc}
\toprule
\multirow{2}{*}{} & \multirow{2}{*}{} & \multicolumn{4}{c}{Training set probe overlap} \\ \cmidrule{3-6} 
 & & 20\% & 38\% & 44\% & 50\% \\ \midrule
\multirow{8}{*}{\rotatebox{90}{Test set probe overlap}} & 20\% & \makecell{\textbf{0.135$\pm$0.026} \\ \textbf{0.854$\pm$0.027}} & \makecell{0.186$\pm$0.030 \\ 0.762$\pm$0.032} & \makecell{0.183$\pm$0.028 \\ 0.744$\pm$0.038} & \makecell{0.200$\pm$0.039 \\ 0.700$\pm$0.044} \\ \cmidrule{2-6} 
 & 38\% & \makecell{0.119$\pm$0.021 \\ 0.873$\pm$0.029} & \makecell{\textbf{0.085$\pm$0.017} \\ \textbf{0.927$\pm$0.023}} & \makecell{0.099$\pm$0.017 \\ 0.903$\pm$0.026} & \makecell{0.101$\pm$0.015 \\ 0.893$\pm$0.025} \\ \cmidrule{2-6} 
 & 44\% & \makecell{0.110$\pm$0.021 \\ 0.892$\pm$0.024} & \makecell{0.087$\pm$0.018 \\ 0.930$\pm$0.024} & \makecell{\textbf{0.078$\pm$0.020} \\ \textbf{0.947$\pm$0.022}} & \makecell{0.080$\pm$0.018 \\ 0.937$\pm$0.022} \\ \cmidrule{2-6} 
 & 50\% & \makecell{0.104$\pm$0.020 \\ 0.906$\pm$0.021} & \makecell{0.079$\pm$0.021 \\ 0.948$\pm$0.020} & \makecell{0.072$\pm$0.020 \\ 0.959$\pm$0.029} & \makecell{\textbf{0.067$\pm$0.020} \\ \textbf{0.967$\pm$0.015}} \\ \bottomrule
\end{tabular}
}
\vspace{-10pt}
\end{table}

\vspace{-10pt}
\section{Discussion}
\vspace{-5pt}
This study introduces a diffusion model-based approach for ptychographic image reconstruction, demonstrating superior performance in low-overlap scenarios. The proposed method achieves an average NRMSE below 0.1 for overlap ratios as low as 38\% and remains competitive even at 20\%, significantly outperforming conventional iterative techniques in such conditions. By enabling accurate reconstructions with reduced probe overlap, this approach has the potential to decrease imaging times, optimize synchrotron beamtime utilization, and minimize radiation exposure for dose-sensitive samples. Notably, the proposed method requires only a single training phase on the target object distribution and can generalize effectively across varying probe overlaps and scanning positions. As for the computational complexity, all methods in this study execute 10,000 evaluations of the forward and adjoint imaging operators. However, the diffusion model-based approach requires additional gradient computations and neural network evaluations at each step, increasing computational complexity. Future work will focus on refining likelihood guidance to reduce the number of diffusion steps and on experimental validations to confirm real-world performance.

\vspace{-10pt}
\bibliographystyle{IEEEbib}
\bibliography{refs, bicer}

\begin{thebibliography}{10}

\bibitem{pfeiffer2018x}
Franz Pfeiffer,
\newblock ``X-ray ptychography,''
\newblock {\em Nature Photonics}, vol. 12, no. 1, pp. 9--17, 2018.

\bibitem{APS_comp_strategy}
``A{P}{S}, {S}cientific {C}omputing {S}trategy document,''
  \url{https://www-dev.aps.anl.gov/sites/www.aps.anl.gov/files/APS-Uploads/XSD/XSD-Strategic-Plans/APSScientificComputingStrategy-2022-12-09-FINAL.pdf},
\newblock [Accessed: Sept. 2024].

\bibitem{benmore2022advancing}
Chris Benmore, Tekin Bicer, Maria~KY Chan, Zichao Di, Do{\u{g}}a G{\"u}rsoy,
  Inhui Hwang, Nikita Kuklev, Dergan Lin, Zhengchun Liu, Ihar Lobach, et~al.,
\newblock ``Advancing ai/ml at the advanced photon source,''
\newblock {\em Synchrotron Radiation News}, vol. 35, no. 4, pp. 28--35, 2022.

\bibitem{zhao2024suppressing}
Chen Zhao, Chuanwei Wang, Xiang Liu, Inhui Hwang, Tianyi Li, Xinwei Zhou,
  Jiecheng Diao, Junjing Deng, Yan Qin, Zhenzhen Yang, et~al.,
\newblock ``Suppressing strain propagation in ultrahigh-ni cathodes during fast
  charging via epitaxial entropy-assisted coating,''
\newblock {\em Nature Energy}, pp. 1--12, 2024.

\bibitem{babu2023deep}
Anakha~V Babu, Tao Zhou, Saugat Kandel, Tekin Bicer, Zhengchun Liu, William
  Judge, Daniel~J Ching, Yi~Jiang, Sinisa Veseli, Steven Henke, et~al.,
\newblock ``Deep learning at the edge enables real-time streaming ptychographic
  imaging,''
\newblock {\em Nature Communications}, vol. 14, no. 1, pp. 7059, 2023.

\bibitem{yu2022scalable}
Xiaodong Yu, Viktor Nikitin, Daniel~J Ching, Selin Aslan, Do{\u{g}}a
  G{\"u}rsoy, and Tekin Bi{\c{c}}er,
\newblock ``Scalable and accurate multi-gpu-based image reconstruction of
  large-scale ptychography data,''
\newblock {\em Scientific reports}, vol. 12, no. 1, pp. 5334, 2022.

\bibitem{yu2021topology}
Xiaodong Yu, Tekin Bicer, Rajkumar Kettimuthu, and Ian Foster,
\newblock ``Topology-aware optimizations for multi-gpu ptychographic image
  reconstruction,''
\newblock in {\em Proceedings of the ACM International Conference on
  Supercomputing}, 2021, pp. 354--366.

\bibitem{aps-raven}
{Intelligence Advanced Research Projects Activity},
\newblock ``{Rapid Analysis of Various Emerging Nanoelectronics},''
  \url{https://www.iarpa.gov/index.php/research-programs/raven},
\newblock [Accessed: April 2020].

\bibitem{gan2024ptychodv}
Weijie Gan, Qiuchen Zhai, Michael~Thompson McCann, Cristina~Garcia Cardona,
  Ulugbek~S Kamilov, and Brendt Wohlberg,
\newblock ``Ptychodv: Vision transformer-based deep unrolling network for
  ptychographic image reconstruction,''
\newblock {\em IEEE Open Journal of Signal Processing}, 2024.

\bibitem{zhai2023projected}
Qiuchen Zhai, Gregery~T Buzzard, Kevin Mertes, Brendt Wohlberg, and Charles~A
  Bouman,
\newblock ``Projected multi-agent consensus equilibrium (pmace) with
  application to ptychography,''
\newblock {\em IEEE Transactions on Computational Imaging}, 2023.

\bibitem{xu2018accelerated}
Rui Xu, Mahdi Soltanolkotabi, Justin~P Haldar, Walter Unglaub, Joshua Zusman,
  Anthony~FJ Levi, and Richard~M Leahy,
\newblock ``Accelerated wirtinger flow: A fast algorithm for ptychography,''
\newblock {\em arXiv preprint arXiv:1806.05546}, 2018.

\bibitem{maiden2017further}
Andrew Maiden, Daniel Johnson, and Peng Li,
\newblock ``Further improvements to the ptychographical iterative engine,''
\newblock {\em Optica}, vol. 4, no. 7, pp. 736--745, 2017.

\bibitem{candes2015phase}
Emmanuel~J Candes, Xiaodong Li, and Mahdi Soltanolkotabi,
\newblock ``Phase retrieval via wirtinger flow: Theory and algorithms,''
\newblock {\em IEEE Transactions on Information Theory}, vol. 61, no. 4, pp.
  1985--2007, 2015.

\bibitem{guan2019ptychonet}
Ziqiao Guan, Esther~H Tsai, Xiaojing Huang, Kevin~G Yager, and Hong Qin,
\newblock ``Ptychonet: Fast and high quality phase retrieval for
  ptychography,''
\newblock Tech. {R}ep., Brookhaven National Lab.(BNL), Upton, NY (United
  States), 2019.

\bibitem{cherukara2020ai}
Mathew~J Cherukara, Tao Zhou, Youssef Nashed, Pablo Enfedaque, Alex Hexemer,
  Ross~J Harder, and Martin~V Holt,
\newblock ``Ai-enabled high-resolution scanning coherent diffraction imaging,''
\newblock {\em Applied Physics Letters}, vol. 117, no. 4, 2020.

\bibitem{barutcu2022compressive}
Semih Barutcu, Do{\u{g}}a G{\"u}rsoy, and Aggelos~K Katsaggelos,
\newblock ``Compressive ptychography using deep image and generative priors,''
\newblock {\em arXiv preprint arXiv:2205.02397}, 2022.

\bibitem{ekmekci2024integrating}
Canberk Ekmekci, Tekin Bicer, Zichao~Wendy Di, Junjing Deng, and Mujdat Cetin,
\newblock ``Integrating generative and physics-based models for ptychographic
  imaging with uncertainty quantification,''
\newblock {\em arXiv preprint arXiv:2412.10882}, 2024.

\bibitem{dhariwal2021diffusion}
Prafulla Dhariwal and Alexander Nichol,
\newblock ``Diffusion models beat gans on image synthesis,''
\newblock {\em Advances in neural information processing systems}, vol. 34, pp.
  8780--8794, 2021.

\bibitem{chung2022diffusion}
Hyungjin Chung, Jeongsol Kim, Michael~T Mccann, Marc~L Klasky, and Jong~Chul
  Ye,
\newblock ``Diffusion posterior sampling for general noisy inverse problems,''
\newblock {\em arXiv preprint arXiv:2209.14687}, 2022.

\bibitem{lugmayr2022repaint}
Andreas Lugmayr, Martin Danelljan, Andres Romero, Fisher Yu, Radu Timofte, and
  Luc Van~Gool,
\newblock ``Repaint: Inpainting using denoising diffusion probabilistic
  models,''
\newblock in {\em Proceedings of the IEEE/CVF conference on computer vision and
  pattern recognition}, 2022, pp. 11461--11471.

\bibitem{zheng2021concept}
Guoan Zheng, Cheng Shen, Shaowei Jiang, Pengming Song, and Changhuei Yang,
\newblock ``Concept, implementations and applications of fourier
  ptychography,''
\newblock {\em Nature Reviews Physics}, vol. 3, no. 3, pp. 207--223, 2021.

\bibitem{ronneberger2015u}
Olaf Ronneberger, Philipp Fischer, and Thomas Brox,
\newblock ``U-net: Convolutional networks for biomedical image segmentation,''
\newblock in {\em Medical image computing and computer-assisted
  intervention--MICCAI 2015: 18th international conference, Munich, Germany,
  October 5-9, 2015, proceedings, part III 18}. Springer, 2015, pp. 234--241.

\bibitem{antholzer2019deep}
Stephan Antholzer, Markus Haltmeier, and Johannes Schwab,
\newblock ``Deep learning for photoacoustic tomography from sparse data,''
\newblock {\em Inverse problems in science and engineering}, vol. 27, no. 7,
  pp. 987--1005, 2019.

\end{thebibliography}

\end{document}